\documentclass[aps,prb,showpacs,preprint]{revtex4-1}%
\usepackage{amsfonts}
\usepackage{amsmath}
\usepackage{amssymb}
\usepackage{graphicx}%
\begin{document}

\title{Temperature effects on magnetoplasmon spectrum of a weakly modulated graphene monolayer}
\author{M. Tahir$^{1\ast}$ , K. Sabeeh$^{2\dag}$, A. MacKinnon$^{3\ddag}$ }

\begin{abstract}
In this work, we determine the effects of temperature on the magnetoplasmon
spectrum of an electrically modulated graphene monolayer as well as a
two-dimensional electron gas (2DEG). The intra-Landau-band magnetoplasmon
spectrum within the Self Consistent Field (SCF) approach is investigated for
both the aforementioned systems. Results obtained not only exhibit
Shubnikov-de Hass (SdH) oscillations but also commensurability oscillations
(Weiss oscillations). These oscillations are periodic as a function of inverse
magnetic field. We find that both the magnetic oscillations, SdH and Weiss,
have a greater amplitude and are more robust against temperature in graphene
compared to a conventional 2DEG. Furthermore, there is a $\pi$ phase shift
between the magnetoplasmon oscillations in the two systems which can be
attributed to Dirac electrons in graphene acquiring a Berry's phase as they
traverse a closed path in a magnetic field.

\end{abstract}
\date{[Date text]date}
\maketitle

\affiliation{$^{1}$Department of Physics, University of Sargodha, Sargodha, Pakistan}
\affiliation{$^{2}$Department of Physics, Quaid-i-Azam University, Islamabad 45320, Pakistan.}
\affiliation{$^{3}$Department of Physics, Imperial College London, South Kensington campus, London SW7 2AZ, U.K.}


\section{Introduction}

Remarkable progress made in epitaxial crystal growth techniques has led to the
fabrication of novel semiconductor heterostructures. These modern
microstructuring techniques have made possible the realization of a
two-dimensional electron gas (2DEG) system in semiconductor heterostructures.
A 2DEG is a condensed matter system where a number of novel phenomena have
been observed over the years. One such phenomenon was the observation of
commensurability oscillations in physical properties when the 2DEG in the
presence of a magnetic field is subjected to electric modulation. The electric
modulation introduces an additional length scale in the system and the
occurrence of these oscillations, commonly known as Weiss oscillations, is due
to the commensurability of the electron cyclotron diameter at the Fermi energy
and the period of the electric modulation. These oscillations were found to be
periodic in the inverse magnetic field \cite{1,2,3}. This type of electrical
modulation of the 2D system can be carried out by depositing an array of
parallel metallic strips on the surface or through two interfering laser beams
\cite{4,5}.

One of the important electronic properties of such a system is the collective
excitations (plasmons). Weiss oscillations in the magnetoplasmon spectrum of a
2DEG has been the subject of continued interest \cite{6,7,8,9}. Recently, the
fabrication of crystalline graphene monolayers has generated a lot of
interest. The study of this new material is not only of academic interest but
there are serious efforts underway to investigate whether Graphene can serve
as the basic material for a carbon-based electronics that might replace
silicon. Graphene has a honeycomb lattice of carbon atoms. The quasiparticles
in graphene have a band structure in which electron and hole bands touch at
two points in the Brillouin zone. At these Dirac points the quasiparticles
obey the massless Dirac equation. In other words, they behave as massless
Dirac fermions leading to a linear dispersion relation $\epsilon_{k}=vk$ (
with the characteristic velocity $v\simeq10^{6}m/s)$. This difference between
the nature of the quasiparticles in graphene and those in a conventional 2DEG
has given rise to a host of new and unusual phenomena, such as anomalous
quantum Hall effects and a $\pi$ Berry's phase\cite{10,11,12,13,14,15}. These
transport experiments have shown results in agreement with the presence of
Dirac fermions. The 2D Dirac-like spectrum was confirmed recently by cyclotron
resonance measurements and also by angle resolved photoelectron spectroscopy
(ARPES) measurements in monolayer graphene\cite{16,17}. Recent theoretical
work on graphene multilayers has also shown the existence of Dirac electrons
with a linear energy spectrum in monolayer graphene\cite{18}. Plasmons in
graphene were studied as early as the eighties\cite{19} and more recently
\cite{20,21,22,23}. In this work, we investigate magnetoplasmons in a graphene
monolayer when it is subjected to electric modulation at finite temperature.
In a 2DEG system, as well as in graphene, plasmons emerge as a result of
electron-electron interactions. In the presence of a magnetic field, these
magnetoplasmons occur at frequencies that oscillate with the magnetic field.
When this system is subjected to an electric modulation, broadening of the
Landau levels occurs resulting in both inter- and intra- Landau band
magnetoplasmons. The former arise as a result of electronic transitions
between different Landau bands, whereas the latter are due to transitions
within a single Landau band. A study of magnetoplasmons at zero temperature
was carried out by us in earlier work \cite{24} where we showed that the
intra-Landau band magnetoplasmons exhibit Weiss oscillations as a result of
modulation. Presently, we extend our earlier work by taking into account the
effects of temperature on the magnetoplasmons in graphene. The motivation for
the finite temperature calculation presented here is that it allows us to
account for SdH oscillations in addition to Weiss oscillations in the
magnetoplasmon spectrum and the effects of temperature on them. While working
on the finite temperature dispersion relation for magnetoplasmons in graphene
we realized that no finite temperature calculation for a 2DEG\cite{9} is
available in the literature. In this work, therefore, we also determine the
finite temperature dispersion relation for magnetoplasmons in an electrically
modulated\cite{9} 2DEG. Furthermore, asymptotic expressions are derived that
allow us to show that the magnetoplasmons in graphene are more robust against
temperature compared to those occurring in a conventional 2DEG realized in
semiconductor heterostructures. In addition, the characteristic temperatures
for damping of the magnetic oscillations (SdH and Weiss) are determined for
both the graphene and 2DEG systems. The experimental study of these should be
as revealing as the enhanced magnetoconductivity prediction\cite{25} and even
more interesting, since it bears directly on the many-body properties of the
two-dimensional systems, as well as frequency dependent transport and optical
response properties.

\section{Temperature effects on the magnetoplasmon spectrum of weakly
modulated graphene}

We consider two-dimensional Dirac electrons in graphene moving in the
$(x-y)$-plane. The magnetic field ($B$) is applied along the $z$-direction
perpendicular to the graphene plane. This system is subjected to weak electric
modulation along the $x$-direction. We employ the Landau gauge and write the
vector potential as $\mathbf{A}=(0,Bx,0)$. Employing the tight-binding
approach it has been shown that the Dirac electrons, in the absence of the
periodic potential and the magnetic field, obey the two dimensional Dirac-like
Hamiltonian \cite{26,27}. The effect of the magnetic field can be included
through the minimal coupling with the result that the two-dimensional, single
electron, Dirac-like Hamiltonian in the Landau gauge is \cite{12,25,26,27,28}%
\begin{equation}
H_{0}=v\sigma.(-i\hslash\mathbf{\nabla}+e\mathbf{A}). \label{1}%
\end{equation}
Here\ $\sigma=\{\sigma_{x},\sigma_{y}\}$ are the Pauli matrices and $v$
characterizes the electron velocity. The complete Hamiltonian of our system is
represented as%
\begin{equation}
H=H_{0}+V(x) \label{2}%
\end{equation}
where $H_{0}$ is the unmodulated Hamiltonian and $V(x)$ is the one-dimensional
periodic modulation potential along the $x$-direction modelled as%
\begin{equation}
V(x)=V_{0}\cos(Kx) \label{3}%
\end{equation}
where $K=2\pi/a,a$ is the period of modulation and $V_{0}$ is the constant
modulation amplitude. Applying standard perturbation theory to determine the
first order correction to the unmodulated energy eigenvalues in the presence
of modulation, we obtain
\begin{equation}
\varepsilon(n,x_{0})=\hslash\omega_{g}\sqrt{n}+F_{n}\cos(Kx_{0}) \label{4}%
\end{equation}
where $\omega_{g}=v\sqrt{2eB/\hslash}$ is the cyclotron frequency of Dirac
electrons in graphene, $n$ is an integer, $F_{n}=\frac{1}{2}V_{0}e^{-\chi
/2}[L_{n}(\chi)+L_{n-1}(\chi)]$, $\chi=\frac{1}{2}K^{2}l^{2}$, $x_{0}%
=l^{2}k_{y}$, $L_{n}(\chi)$ are Laguerre polynomials and $l=\sqrt{\hslash/eB}$
is the magnetic length. The Landau level spectrum for Dirac electrons in
graphene ($\varepsilon_{n}=\sqrt{n}\hslash\omega_{g}$) is significantly
different from the spectrum for electrons in a conventional 2DEG, which is
given as $\varepsilon(n)=\hbar\omega_{c}(n+1/2)$, where $\omega_{c}$ is the
cyclotron frequency.

The intra-Landau-band plasmon spectrum is determined within the
Ehrenreich-Cohen self-consistent-field (SCF) approach\cite{29}. Following
reference~\cite{24} the plasmon dispersion relation is given by solving
\begin{equation}
1=\frac{2\pi e^{2}}{k\bar{q}}\frac{1}{\pi al^{2}}\sum_{n,n^{\prime}%
}C_{nn^{\prime}}\left(  \frac{\bar{q}^{2}}{2eB}\right)  \left[  I_{1}%
(\omega)+I_{1}(-\omega)\right]  \label{eq:4a}%
\end{equation}
for $\omega$, where
\begin{equation}
I_{1}(\omega)=P\int_{0}^{a}\mathrm{d}x_{0}\frac{f\left[  \varepsilon
(n,x_{0})\right]  }{\varepsilon(n,x_{0})-\varepsilon(n,x_{0}+x_{0}^{\prime
})+\omega}\,, \label{eq:4b}%
\end{equation}
$x_{0}^{\prime}=l^{2}q_{y},\bar{q}$ is the two dimensional wave number, $k$ is
background dielectric constant and $f\left[  \varepsilon(n,x_{0})\right]
$\ is the Fermi distribution function. This result was obtained previously in
\cite{24} where it was used to obtain the zero temperature dispersion
relation. Here we are interested in temperature effects to account for SdH
oscillations, in addition to Weiss oscillations, in the magnetoplasmon
spectrum. We proceed as follows to determine the finite temperature dispersion
relation. In the regime of weak modulation, the distribution function
$f\left[  \varepsilon(n,x_{0})\right]  $ can be expressed as%
\begin{equation}
f\left[  \varepsilon(n,x_{0})\right]  \simeq f(\varepsilon_{n})+F_{n}%
f^{\prime}(\varepsilon_{n})\cos(Kx_{0}), \label{7}%
\end{equation}
where $f^{\prime}(x)=\frac{d}{dx}f(x)$. Substituting the above expression for
$f\left[  \varepsilon(n,x_{0})\right]  $, the integral over $x_{0}$ may be
written as
\begin{align}
I_{1}(\omega)  &  =K^{-1}\int_{0}^{2\pi}\left[  \frac{f(\varepsilon
_{n})+f^{\prime}(\varepsilon_{n})F_{n}\sin({\textstyle\frac{1}{2}}%
Kx_{0}^{\prime})\sin\theta}{2F_{n}\sin({\textstyle\frac{1}{2}}Kx_{0}^{\prime
})\sin\theta+\omega}\right]  \mathrm{d}\theta\label{eq:7b}\\
&  =K^{-1}f(\varepsilon_{n})\int_{0}^{2\pi}\frac{\mathrm{d}\theta}{\Phi
_{n}\sin\theta+\omega}+K^{-1}f^{\prime}(\varepsilon_{n}){\textstyle\frac{1}%
{2}}\Phi_{n}\int_{0}^{2\pi}\frac{\sin\theta\,\mathrm{d}\theta}{\Phi_{n}%
\sin\theta+\omega}\,, \label{eq:7c}%
\end{align}
where $\Phi_{n}=2F_{n}\sin({\textstyle\frac{1}{2}}Kx_{0}^{\prime})$. There is
also a term with cosines in the numerator, but the resulting integral is
easily shown to be zero by symmetry around $\theta=\textstyle{\frac{\pi}{2}}$.
The first integral in (\ref{eq:7c}) can be solved by rewriting it in terms of
a contour integral around a unit circle to obtain
\begin{equation}
\int_{0}^{2\pi}\frac{\mathrm{d}\theta}{\Phi_{n}\sin\theta+\omega}=\left\{
\begin{array}
[c]{ll}%
\displaystyle+\frac{2\pi}{\sqrt{\omega^{2}-\Phi_{n}^{2}}} &
\displaystyle\mbox{ if }\frac{\omega}{\Phi_{n}}>1\\
\displaystyle\pm\frac{2\pi\mathrm{i}}{\sqrt{\Phi_{n}^{2}-\omega^{2}}} &
\displaystyle\mbox{ if }-1<\frac{\omega}{\Phi_{n}}<1\\
\displaystyle-\frac{2\pi}{\sqrt{\omega^{2}-\Phi_{n}^{2}}} &
\displaystyle\mbox{ if }\frac{\omega}{\Phi_{n}}<-1\,,\label{eq:7d}%
\end{array}
\right.
\end{equation}
as long as $\omega$ has an infinitesimal but non-zero imaginary part, such
that one pole always remains inside and one outside the unit circle. If
$\omega$ must be real, however, both poles may be on the unit circle and their
contributions cancel to leave a principal value of zero. Either way we deduce
that our problem has no real finite solutions for small $\omega$. In fact, it
has no solution unless $\omega$ is larger than all of the $\Phi_{n}$. However,
this is not the integral we need as it will cancel in the expression
$I(\omega)+I(-\omega)$. The 2nd integral in (\ref{eq:7c}) may be deduced from
(\ref{eq:7d}) using
\begin{align}
\Phi_{n}\int_{0}^{2\pi}\frac{\sin\theta\,\mathrm{d}\theta}{\Phi_{n}\sin
\theta+\omega}  &  =2\pi-\omega\int_{0}^{2\pi}\frac{\mathrm{d}\theta}{\Phi
_{n}\sin\theta+\omega}\\
&  =2\pi\left[  1-\frac{\omega}{\sqrt{\omega^{2}-\Phi_{n}^{2}}}%
\mathop{\mbox{sign}}\left(  \frac{\omega}{\Phi_{n}}\right)  \right]  \,,
\label{eq:7e}%
\end{align}
which reduces to $2\pi$ for $\omega^{2}<\Phi_{n}^{2}$ if we use the principal
value interpretation of (\ref{eq:7d}). We note that this expression may be
expanded in the form
\begin{equation}
\Phi_{n}\int_{0}^{2\pi}\frac{\sin\theta\,\mathrm{d}\theta}{\Phi_{n}\sin
\theta+\omega}=-\pi\frac{\Phi_{n}^{2}}{\omega^{2}}+\mathop{\mathrm{O}}\left[
\left(  \frac{\Phi_{n}}{\omega}\right)  ^{4}\right]  \label{eq:7f}%
\end{equation}
for $\omega^{2} >\Phi_{n}^{2}$.

Substituting this back into (\ref{eq:4a}) and concentrating on intra-Landau
level plasmons by setting $C_{nn^{\prime}}=\delta_{nn^{\prime}}$ we obtain
\begin{align}
1  &  =\frac{2\pi e^{2}}{k\bar{q}}\frac{1}{\pi al^{2}}K^{-1}\sum_{n}f^{\prime
}(\varepsilon_{n})2\pi\left[  1-\frac{\omega}{\sqrt{\omega^{2}-\Phi_{n}^{2}}%
}\mathop{\mbox{sign}}\left(  \frac{\omega}{\Phi_{n}}\right)  \right] \\
&  \approx\frac{2\pi e^{2}}{k\bar{q}}\frac{1}{\pi al^{2}}K^{-1}\sum_{n}\left[
-f^{\prime}(\varepsilon_{n})\right]  \pi\frac{\Phi_{n}^{2}}{\omega^{2}}\,.
\label{eq:7g}%
\end{align}
We note, at this point that $V_{0}$, $F_{n}$ and $\omega$ are of similar
orders of magnitude (meV), but that, in the long wavelength (small $q_{y}$)
limit $x_{0}^{\prime}\rightarrow0$ (recall that $x_{0}^{\prime}=l^{2}q_{y}$),
$\Phi_{n}\ll\omega$. We are therefore justified in taking the leading term in
the expansion in (\ref{eq:7g}), which can be solved to give the intra-Landau
band plasmon excitations frequency ($\tilde\omega$) as
\begin{equation}
\tilde{\omega}^{2}=\frac{4e^{2}}{k\bar{q}l^{2}}\sin^{2}\left(  \frac{\pi
x_{0}^{\prime}}{a}\right)  \times G_{\mathrm{g}}, \label{8}%
\end{equation}
with $G_{\mathrm{g}}=\underset{n}{\sum}F_{n}^{2}\times\lbrack-f^{\prime
}(\varepsilon_{n})]$. This result will be valid except for low frequencies,
where $G_{\mathrm{g}}$ may acquire an $\omega$ dependence and cease to exist
before $\omega=0$. The resulting physics can be made more transparent by
considering the asymptotic expression for the intra-Landau band magnetoplasmon
spectrum, where analytic results can be obtained in terms of elementary functions.

To obtain an expression for the intra-Landau band plasmon spectrum we employ
the following asymptotic expression for the Laguerre polynomials%
\begin{equation}
\exp^{-\chi/2}L_{n}(\chi)\rightarrow\frac{1}{\sqrt{\pi\sqrt{n\chi}}}%
\cos\left(  2\sqrt{n\chi}-\frac{\pi}{4}\right)  \label{9}%
\end{equation}
Note that the asymptotic results are valid when many Landau Levels are filled.
We\ now take the continuum limit:%
\begin{equation}
n\rightarrow\frac{\varepsilon^{2}}{\hslash^{2}\omega_{g}^{2}},\qquad\sum
_{n=0}^{\infty}\rightarrow%
{\displaystyle\int\limits_{0}^{\infty}}
\frac{2\varepsilon\,d\varepsilon}{\hslash^{2}\omega_{g}^{2}}.\label{10}%
\end{equation}
$G_{\mathrm{g}}$, which appears in the expression for $\tilde{\omega}^{2}$ in
equation~(\ref{8}) can be expressed as%
\begin{equation}
G_{\mathrm{g}}=\frac{V_{0}^{2}}{\pi}%
{\displaystyle\int\limits_{0}^{\infty}}
\frac{2\varepsilon\,d\varepsilon}{\hslash^{2}\omega_{g}^{2}\left(
\varepsilon/\hslash\omega_{g}\right)  \sqrt{\chi}}\frac{\beta g(\varepsilon
)}{[g(\varepsilon)+1)]^{2}}\cos^{2}\left(  2\sqrt{n\chi}-\frac{\pi}{4}\right)
\cos^{2}\left(  \frac{1}{2}\sqrt{\frac{\chi}{n}}\right)  \label{11}%
\end{equation}
where $g(\varepsilon)=\exp\left[  \beta(\varepsilon-\varepsilon_{F})\right]
$, $\varepsilon_{F}=\hslash vK_{F},K_{F}=\sqrt{2\pi n_{D}}$, $n_{D}$ is the
number density, $\chi=\frac{1}{2}{K^{2}l^{2}}=2\pi^{2}/b$ with $b={eBa^{2}%
}/{\hslash}$ and $\beta={1}/{K_{B}T}$. Assuming that the temperature is
sufficiently low that $\beta^{-1}\ll\varepsilon_{F}$ and substituting
$\varepsilon=\varepsilon_{F}+s\beta^{-1}$, we rewrite the above integral as%
\begin{equation}
G_{\mathrm{g}}=\frac{2V_{0}^{2}\cos^{2}\left(  \frac{\hslash\omega_{g}%
}{2\varepsilon_{F}}\sqrt{\chi}\right)  }{\pi\hslash\omega_{g}\sqrt{\chi}}%
{\displaystyle\int\limits_{0}^{\infty}}
ds\frac{e^{s}}{[e^{s}+1)]^{2}}\cos^{2}\left(  \frac{2\varepsilon_{F}\sqrt
{\chi}}{\hslash\omega_{g}}-\frac{\pi}{4}+\frac{2\sqrt{\chi}\beta^{-1}}%
{\hslash\omega_{g}}s\right)  \label{12}%
\end{equation}
with the result%
\begin{equation}
G_{\mathrm{g}}=\frac{V_{0}^{2}\cos^{2}\left(  \frac{\hslash\omega_{g}%
}{2\varepsilon_{F}}\sqrt{\chi}\right)  }{2\pi\hslash\omega_{g}\sqrt{\chi}%
}\left[  2-2A\left(  \frac{T}{T_{\mathrm{W}}}\right)  +4A\left(  \frac
{T}{T_{\mathrm{W}}}\right)  \cos^{2}\left(  \frac{2\varepsilon_{F}\sqrt{\chi}%
}{\hslash\omega_{g}}-\frac{\pi}{4}\right)  \right]  \label{13}%
\end{equation}
where
\begin{align*}
A\left(  \frac{T}{T_{\mathrm{W}}}\right)   &  ={\frac{T}{T_{\mathrm{W}}}%
}/{\sinh\left(  \frac{T}{T_{\mathrm{W}}}\right)  }\\
\lim_{\frac{T}{T_{\mathrm{W}}}\rightarrow\infty}A\left(  \frac{T}%
{T_{\mathrm{W}}}\right)   &  =2\frac{T}{T_{\mathrm{W}}}\exp\left(  -\frac
{T}{T_{\mathrm{W}}}\right)  \\
\frac{T}{T_{\mathrm{W}}} &  =\frac{4\pi\sqrt{\chi}K_{B}T}{\hslash\omega_{g}}%
\end{align*}
and $T_{\mathrm{W}}={\hslash\omega_{g}}/{4\pi K_{B}\sqrt{\chi}}$ is the
characteristic damping temperature of the Weiss oscillations. Substituting the
expression for $G_{\mathrm{g}}$ in equation(\ref{8}), the asymptotic
expression for the intra-Landau spectrum is obtained%
\begin{align}
\tilde{\omega}^{2} &  =\frac{2e^{2}}{k\bar{q}\pi l^{2}}\frac{V_{0}^{2}\cos
^{2}\left(  \frac{\omega_{g}}{2\varepsilon_{F}}\sqrt{\chi}\right)  }%
{\pi\hslash\omega_{g}\sqrt{\chi}}\sin^{2}\left(  \frac{\pi}{a}(x_{0}^{\prime
})\right)  \nonumber\\
&  \times\left[  2-2A\left(  \frac{T}{T_{\mathrm{W}}}\right)  +4A\left(
\frac{T}{T_{\mathrm{W}}}\right)  \cos^{2}\left(  \frac{2\varepsilon_{F}%
\sqrt{\chi}}{\hslash\omega_{g}}-\frac{\pi}{4}\right)  \right]  \label{14}%
\end{align}
As the above expression is only valid at high temperature ($K_{B}%
T>>\hslash\omega_{g}/2\pi^{2}$) it is not able to account for the SdH
oscillations occurring in the magnetoplasmon spectrum. To take these into
account we use the following expression for the density of states at low
magnetic fields in the absence of impurity scattering
\[
D(\varepsilon)=\frac{\varepsilon}{\pi l^{2}\hslash^{2}\omega_{g}^{2}}\left[
1+2\cos\left(  \frac{2\pi\varepsilon^{2}}{\hslash^{2}\omega_{g}^{2}}\right)
\right]  \,.
\]
This expression for $D(\varepsilon)$\ was obtained in \cite{30,31} in the
absence of scattering and with gap $\Delta=0$. The sum can now be expressed in
the continuum approximation as $\overset{\infty}{\underset{n=0}{%
{\displaystyle\sum}
}}\mapsto2\pi l^{2}%
{\displaystyle\int\limits_{0}^{\infty}}
D(\varepsilon)d\varepsilon$. Therefore, the intra-Landau band magnetoplasmon
dispersion relation that takes into account both Weiss and SdH oscillations is
given by%
\begin{align}
\tilde{\omega}^{2} &  =\frac{2e^{2}}{k\bar{q}\pi l^{2}}\frac{V_{0}^{2}\cos
^{2}\left(  \frac{\hslash\omega_{g}}{2\varepsilon_{F}}\sqrt{\chi}\right)
}{\pi\hslash\omega_{g}\sqrt{\chi}}\sin^{2}\left(  \frac{\pi}{a}(x_{0}^{\prime
})\right)  \nonumber\\
&  \times\left\{  \left[  2-2A\left(  \frac{T}{T_{\mathrm{W}}}\right)
+4A\left(  \frac{T}{T_{\mathrm{W}}}\right)  \cos^{2}\left(  \frac
{2\varepsilon_{F}\sqrt{\chi}}{\hslash\omega_{g}}-\frac{\pi}{4}\right)
\right]  \right.  \nonumber\\
&  \left.  +4\frac{2\frac{T}{T_{SdH}}}{\sinh(\frac{T}{T_{SdH}})}\cos\left(
\frac{2\pi\varepsilon_{F}^{2}}{\hslash^{2}\omega_{g}^{2}}\right)  \cos
^{2}\left(  \frac{2\varepsilon_{F}\sqrt{\chi}}{\hslash\omega_{g}}-\frac{\pi
}{4}\right)  \right\}  \label{15}%
\end{align}
where%
\begin{align*}
\frac{T}{T_{\mbox{\scriptsize SdH}}} &  =\frac{4\pi^{2}\varepsilon_{F}K_{B}%
T}{\hslash^{2}\omega_{g}^{2}}\\
T_{\mbox{\scriptsize SdH}} &  =\frac{\hslash^{2}\omega_{g}^{2}}{4\pi
^{2}\varepsilon_{F}K_{B}}%
\end{align*}
is the characteristic damping temperature of the SdH oscillations.

Following the same approach as given above for a graphene monolayer, we can
obtain the intra-Landau band magnetoplasmon spectrum for a conventional 2DEG
at finite temperature with the result%
\begin{equation}
\tilde{\omega}^{2}=\frac{4e^{2}m^{\ast}\omega_{c}}{\hslash k\bar{q}\pi}%
\sin^{2}\left[  \frac{\pi}{a}(x_{0}^{\prime})\right]  \times G_{\mathrm{c}%
},\label{16}%
\end{equation}
where $m^{\ast}$ is the standard electron mass, $G_{\mathrm{c}}=\underset
{n}{\sum}F_{n}^{2}(C)\times\left[  -f^{\prime}(\varepsilon(n))\right]  $, and
$F_{n}(C)=V_{0}e^{-\chi/2}L_{n}(\chi)$ is the modulation width of the
conventional 2DEG. The corresponding asymptotic result is%
\begin{align}
\tilde{\omega}^{2} &  =\frac{4V_{0}^{2}e^{2}m^{\ast}\omega_{c}}{k\bar{q}%
2\pi^{2}\hslash\sqrt{\chi\hslash\omega_{c}\varepsilon_{F}}}\sin^{2}\left(
\frac{\pi}{a}(x_{0}^{\prime})\right)  \nonumber\\
&  \times\left\{  \left[  1-A\left(  \frac{T}{T_{\mathrm{W}}^{p}}\right)
+2A\left(  \frac{T}{T_{\mathrm{W}}^{p}}\right)  \cos^{2}\left(  2\sqrt
{\frac{\chi\varepsilon_{F}}{\hslash\omega_{c}}}-\frac{\pi}{4}\right)  \right]
\right.  \nonumber\\
&  \left.  +4\frac{2\frac{T}{T_{SdH}^{p}}}{\sinh(\frac{T}{T_{SdH}^{p}})}%
\cos\left(  \frac{2\pi\varepsilon_{F}}{\hslash\omega_{c}}-\pi\right)  \cos
^{2}\left[  2\sqrt{\frac{\chi\varepsilon_{F}}{\hslash\omega_{c}}}-\frac{\pi
}{4}\right]  \right\}  \label{17}%
\end{align}
where
\begin{align*}
\frac{T}{T_{\mathrm{W}}^{p}} &  =\frac{4\pi^{2}k_{B}T}{\hslash\omega_{c}%
aK_{F}}\\
\frac{T}{T_{\mbox{\scriptsize SdH}}^{p}} &  =\frac{2\pi^{2}k_{B}T}%
{\hslash\omega_{c}}\\
\chi &  =\frac{K^{2}l^{2}}{2}=2\pi^{2}/b\\
b &  =\frac{eBa^{2}}{\hslash}\,.
\end{align*}
Now we can compare the exact results for the temperature dependent
magnetoplasmon in modulated graphene, as given in equation(\ref{15}) in terms
of the elementary functions\cite{32,33}, with that of a conventional 2DEG
derived in equation(\ref{17}). The differences are:

\begin{itemize}
\item the standard electron energy eigenvalues scale linearly with the
magnetic field whereas those for Dirac electrons in graphene sale as the
square root.

\item the temperature dependence of the Weiss oscillations, $A\left(
{T}/{T_{\mathrm{W}}}\right)  $, is clearly different from that of the standard
2DEG, $A\left(  {T}/{T_{\mathrm{W}}^{p}}\right)  $.

\item the temperature dependence of SdH oscillations in graphene, $\frac
{T}{T_{SdH}}/{\sinh(\frac{T}{T_{SdH}})}$ is different from that of standard
2DEG, $\frac{T}{T_{SdH}^{p}}/{\sinh(\frac{T}{T_{SdH}^{p}})}$.

\item the density of states term that contains the cosine function responsible
for the SdH oscillations has a different dependence in each of the systems:
$\cos\left(  {2\pi\varepsilon_{F}^{2}}/{\hslash^{2}\omega_{g}^{2}}\right)  $
and $\cos\left(  {2\pi\varepsilon_{F}}/{\hslash\omega_{c}}-\pi\right)  $ respectively.
\end{itemize}

These differences will give contrasting results for the temperature dependent
magnetoplasmon spectrum, as we discuss in the next section.

\section{Discussion of Results}

\begin{figure}[tbh]
\begin{center}
\includegraphics[width=0.7\textwidth]{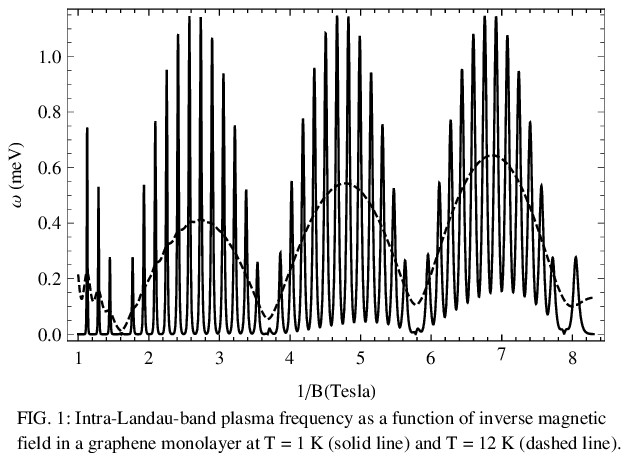}
\end{center}
\end{figure}

The finite temperature intra-Landau band magnetoplasmon dispersion relation
for electrically modulated graphene and 2DEG given by equations (\ref{15}) and
(\ref{17}) are the central results of this work. These results allow us to see
clearly the effects of temperature on the Weiss and SdH oscillations in the
magnetoplasmon spectrum of the two systems. We find that these two types of
oscillations have different characteristic damping temperatures. This is
easily understood if we realize that these oscillations have quite different
origins. The SdH oscillations arise due to the discreteness of the Landau
levels and their observation requires that the thermal energy $K_{B}T$ of the
electrons at temperature $T$ has to be smaller than the separation between the
levels. Weiss oscillations are related to the commensurability of two lengths:
the size of the cyclotron orbit and the period of the modulation. These
oscillations will be observed if the spread in the cyclotron diameter is
smaller than the period of the modulation. To clarify the effects of
temperature we present the magnetoplasmon spectrum in graphene and 2DEG at two
different temperatures in Figs.~(1) and (2).

\begin{figure}[tbh]
\begin{center}
\includegraphics[width=0.7\textwidth]{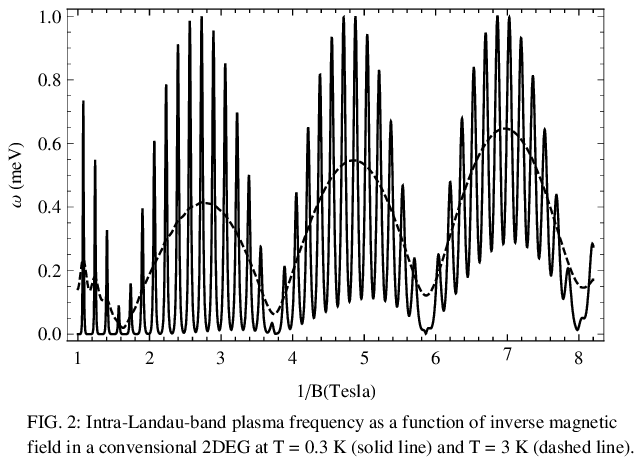}
\end{center}
\end{figure}

In Fig.(1) we show the magnetoplasmon energy as a function of the inverse
magnetic field in a graphene monolayer at two different temperatures: $T=1K$
and $12K.$ The following parameters were employed for
graphene\cite{20,21,22,23,24,25}: $k=3$, $n_{D}=3.16\times10^{16}%
\mathrm{m}^{-2}$, $v=10^{6}\mathrm{m}\mathrm{s}^{-1}$, $a=380\mathrm{nm}$ and
$V_{0}=1.0\mathrm{meV}$. We also take $q_{x}=0$ and $q_{y}=.01k_{F}$, with
$k_{F}=(2\pi n_{D})^{1/2}$. Weiss oscillations superimposed on the SdH
oscillations can be clearly seen in both the curves. In Eq.(\ref{15}) the
terms containing the characteristic temperature for Weiss oscillations
$T_{\mathrm{W}}$ are mainly responsible for Weiss oscillations whereas terms
containing $T_{\mbox{\scriptsize SdH}}$ are responsible for the SdH
oscillations. The intra-Landau-band plasmons have frequencies comparable to
the bandwidth and they occur as a result of the broadening of the Landau
levels due to the modulation in the system. This type of intra-Landau-band
plasmons accompanied by regular oscillatory behavior (in $1/B)$ of the SdH
type was first predicted in \cite{34,35} for a tunneling planar superlattice
where the overlap of the electron wavefunction in adjacent quantum wells
provides the mechanism for the broadening of the Landau levels. The SdH
oscillations occur as a result of the emptying of electrons from successive
Landau levels when they pass through the Fermi level as the magnetic field is
increased. The amplitude of these oscillations is a monotonic function of the
magnetic field when the Landau bandwidth is independent of the band index $n$.
In the density modulated case, the Landau bandwidths oscillate as a function
of the band index $n$, with the result that, in the plasmon spectrum of the
intra-Landau band type, there is a new kind of oscillation named Weiss
oscillation which is also periodic in $1/B$ but with a different period and
amplitude from the SdH type oscillation. At $T=12K$ we find that the SdH
oscillations are washed out while the Weiss oscillations persist. From
Eq.(\ref{15}) we see that $T_{\mathrm{W}}$ and $T_{\mbox{\scriptsize SdH}}$
set the temperatures at which these oscillations will be damped. For the 2DEG
the magnetoplasmon energy as a function of inverse magnetic field is presented
in Fig.(2) at two different temperatures: $T=0.3\mathrm{K}$ and $3\mathrm{K}$.
For a conventional 2DEG (a 2DEG at a GaAs-AlGaAs heterojunction) we use the
following parameters\cite{1,2,3,4,5,6,7,8,9}: $m^{\ast}=.07m_{e}$ ($m_{e}$ is
the electron mass), $k=12$ and $n_{D}=3.16\times10^{15}\mathrm{m}^{-2}$ with
the modulation strength and the period as in the graphene system. In this case
we find that the SdH oscillations in the magnetoplasmon spectrum die out at
$T=3\mathrm{K}$, while they are present at a higher temperature ($12\mathrm{K}%
$) in graphene.

\begin{figure}[tbh]
\begin{center}
\includegraphics[width=0.7\textwidth]{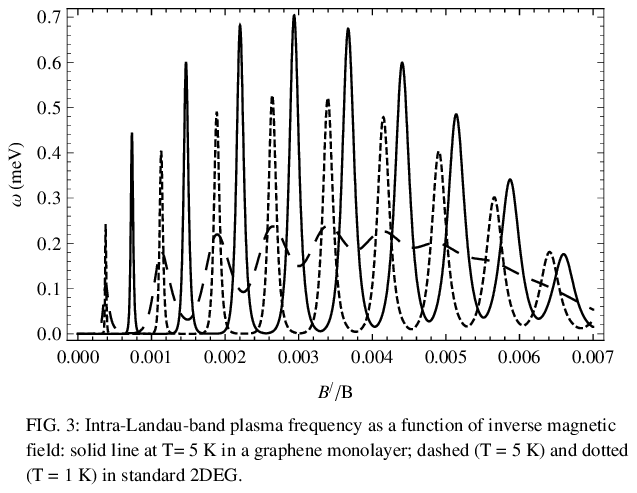}
\end{center}
\end{figure}

To compare the results for the two systems we show in Fig.(3) the
magnetoplasmon spectrum as a function of inverse magnetic field ($\frac
{B^{\prime}}{B}=\frac{\hslash}{Bea^{2}}$, where $B^{\prime}=\frac{\hslash
}{ea^{2}}=0.0046\mbox{Tesla}$) for both graphene (solid line at temperature
$5\mathrm{K}$) and the 2DEG (dotted and dashed line at temperature
$1\mathrm{K}$ and $5\mathrm{K}$ respectively). The oscillations in the
conventional 2DEG have been damped out strongly at $5\mathrm{K}$ but these are
well resolved, significant and have large amplitude in graphene at the same
temperature. This confirms that graphene oscillations are robust and enhanced
against temperature compared to those in a conventional 2DEG. The magnetic
oscillations, SdH and Weiss, have a greater amplitude in graphene compared to
a 2DEG. This can be attributed to the larger characteristic velocity,
$v\sim10^{6}\mathrm{m}\mathrm{s}^{-1}$ of electrons in graphene compared to
the Fermi velocity of standard electrons as well as the smaller background
dielectric constant $k$ in graphene compared to a conventional 2DEG. The
temperature at which we expect the Weiss oscillations to dampen is determined
by comparing the characteristic temperatures for Weiss oscillations in the two
systems: ${T_{\mathrm{W}}^{p}}/{T_{\mathrm{W}}}={v_{F}}/{v}$. The ratio of the
characteristic temperatures is equal to that of the corresponding velocities
at the Fermi surface\cite{25}. For the parameters used in this work
${T_{\mathrm{W}}^{p}}/{T_{\mathrm{W}}}\sim0.24$, implying that Weiss
oscillations are damped at a higher temperature in graphene compared to the
2DEG. Similarly, the damping of SdH oscillations can also be compared in the
two systems through their corresponding characteristic temperatures:
${T_{\mbox{\scriptsize SdH}}^{p}}/{T_{\mbox{\scriptsize SdH}}}={\hslash K_{F}%
}/{vm^{\ast}}\sim0.23$ for $n_{D}=3.16\times10^{15}\mathrm{m}^{-2}$ whereas it
is $\sim0.74$ for $n_{D}=3.16\times10^{16}\mathrm{m}^{-2}$. Hence SdH
oscillations are damped at a higher temperature in graphene compared to a
2DEG. Furthermore, our results show that these oscillations in the
magnetoplasmon spectrum differ in phase by $\pi$ in the two systems, which is
due to quasiparticles in graphene acquiring a Berry's phase of $\pi$ as they
move in the magnetic field\cite{12,13,14,15}.

\section{Conclusions}

We have determined the finite temperature intra-Landau band magnetoplasmon
frequency for both electrically modulated graphene and for a 2DEG in the
presence of a magnetic field, by employing the SCF approach. We find that the
magnetic oscillations (SdH and Weiss) in the magnetoplasmon spectrum in a
graphene monolayer have a higher amplitude compared to the conventional 2DEG
realized in semiconductor heterostructures. Moreover, these oscillations
persist at a higher temperature in graphene compared to the 2DEG. Hence they
are more robust against temperature in graphene. Furthermore, the $\pi$
Berry's phase acquired by the Dirac electrons leads to $\pi$ phase shift in
the magnetoplasmon spectrum in a graphene monolayer compared to a 2DEG.

One of us (K. Sabeeh) would like to acknowledge the support of the Higher
Education Commission (HEC) of Pakistan through project No. 20-1484/R\&D/09. M.
Tahir would like to acknowledge the support of the Pakistan Higher Education
Commission (HEC).

$\ast$ E-mail: m.tahir@uos.edu.pk

$\dag$ E-mail: ksabeeh@qau.edu.pk, kashifsabeeh@hotmail.com

$\ddag$ E-Mail: a.mackinnon@imperial.ac.uk

\end{document}